\documentclass[a4paper]{jpconf}
\usepackage{graphicx}
\usepackage[square,sort&compress]{natbib}

\begin{document}
\title{Optical Integral Field Spectroscopy of NGC 5850}

\author{M Bremer$^1$, J Scharw\"achter$^2$, A Eckart$^{1,3}$, J Zuther$^{1}$, S Fischer$^{1}$, \\
M Valencia-S.$^{1,3}$}

\address{$^1$I.Physikalisches Institut, Universit\"at zu K\"oln,
              Z\"ulpicher Str.77, 50937 K\"oln, Germany\\
         $^2$Research School of Astronomy and Astrophysics, The Australian National University, Mount Stromlo Observatory, Cotter Road, Weston Creek 2611, Australia\\
         $^3$Max-Planck-Institut f\"ur Radioastronomie, Auf dem H\"ugel 69, 53121 Bonn, Germany\\
             }

\ead{mbremer@ph1.uni-koeln.de}

\begin{abstract}
Here we present the preliminary results of the analysis of VIMOS observations of the central 4.5 kpc of the double-barred galaxy NGC 5850. We use optical diagnostic diagrams to study the main ionization mechanism across the field of view confirming the LINER nature in the continuum peak location. Also a star-forming (SF) region is found close to it ($0.46\mathrm{ kpc}$), a second SF region is located east of the center ($1.6\mathrm{kpc}$). Further the data reveals complex nuclear gas kinematics which is likely to be dominated by the secondary bar.
\end{abstract}

\section{Introduction}

Scaling relations, like the one between the mass of the central black hole and the stellar velocity dispersion of the bulge, suggest the co-evolution of the central engine and its host; or at least some form of feedback between them.\\

NGC 5850 is a prototype double-barred early-type spiral SB(r) b galaxy. It is located at a distance of $34.2$ Mpc \cite{Wozniak95} and hosts a polar gaseous disk \cite{doublebarred_structure_moiseev}. Its two spiral arms form ring-like structure, almost resembling the morphology of the center, where the inner bar is surrounded by a (pseudo-) ring. The north-western spiral arm is distorted probably due to a recent ($< 200$ Myr) high velocity encounter with NGC 5846 \cite{Higdon_encounter}, which is $10'$ to the north-west of NGC 5850. The low X-ray luminosity $L_X(0.5-3\ \mathrm{keV})=10^{40.36}\ \mathrm{erg\ s^{-1}}$ \cite{5850_Lx} suggests NGC 5850 to be in a state of low activity. The high luminosity of [NII] lines at the center classifies it as a Low Ionisation Nuclear Emission Region (LINER) \cite{Heckman_Liner, Liner_reference}. Proposed by \cite{Shlosman_bar_in_bar} and simulated by \cite{Friedli_bar_in_bar}, nested bars can be a mechanism to fuel AGN. Its morphology and proximity make NGC 5850 an ideal object to study fuelling and feedback processes. Integral Field Spectroscopy (IFS) in the visible wavelength regime allows us to study in detail the excitation mechanisms of the gas and investigate gas kinematics.

\section{Observations and Data Reduction}

The observations were performed with the integral field spectrometer VIMOS on the Very Large Telescope (VLT) of the European Southern Observatory (ESO) and cover the wavelength range from $4100\ -\ 8800\ $\AA $\ $ with blue and red gratings at high spectral and spatial resolution ($R\approx2300\ \mathrm{at}\ \lambda=5500\ \mathrm{\AA}$, $R\approx3200\ \mathrm{at}\ \lambda=7500\ \mathrm{\AA}$, $0.67"/px \approx 0.163\ \mathrm{kpc}$, $\mathrm{FWHM(PSF)} \approx 2\ \mathrm{px}$). The source was observed in four nights in 2009, two images per night and per grating. The images were reduced with the VIMOS pipeline provided by ESO. We additionally applied a fringing correction. The resulting cubes were corrected for pixel-to-pixel variation and subsequently averaged for each grating separately. Flux calibrations were done by the VIMOS pipeline, but due to a number of bad pixels in the image of the standard star, the values are not reliable yet. However, the absolute flux calibration is not relevant for the preliminary results presented here, but will be improved for future analysis purposes. A sky correction was not performed, because no sky emission line lies close to those of the galaxy and the sky continuum contribution is neglible. \\

After correcting for intergalactic extinction we shifted the spectra to the rest frame. The stellar continuum was subtracted by application of the STARLIGHT program by \cite{starlight} combined with the theoretical models by Bruzual and Charlot \cite{BC03}. The emission lines were fitted in IDL with simple Gaussians using the MPFITPEAK and MPFITEXP routines. The H$\alpha$ and the [NII] $\lambda 6549$ lines were fitted together, though they are not blended, as well as the [SII], $\lambda\lambda6718+6732$ lines. The latter were tied to each other by the constant wavelength difference between their peak values.

\section{Preliminary Results}
\subsection{Emission Maps and Diagnostics}

\begin{figure}
\begin{center}
\includegraphics{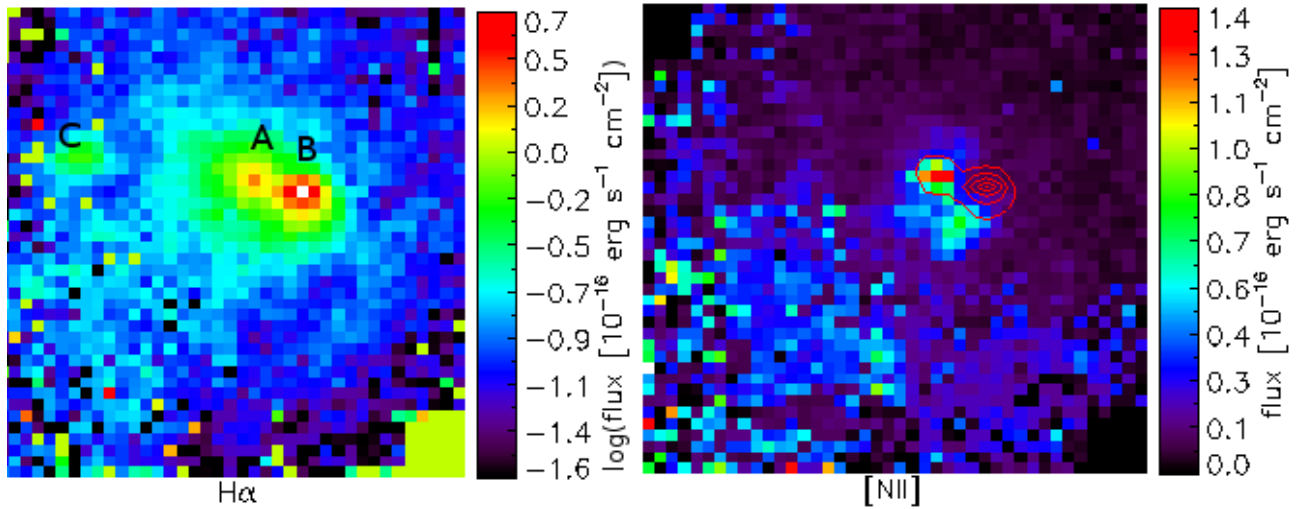}
\end{center}
\caption{\label{H_alpha_NII_linemap}$27"\times27"$ emission maps; north is up, east is left. Left: Line map of H$\alpha$ emission. The location of the continuum peak is labeled with 'A', the HII regions are labeled with 'B' and 'C'. In this and in all other images $1\ \mathrm{px}$ corresponds to approximately $0.163\ \mathrm{kpc}$. Right: Line map of [NII] $\lambda6585$ emission with overplotted contour lines for $95,\ 90,\ 80,\ 60$ and $40\%$ of the H$\alpha$ emission.}
\end{figure}

\begin{figure}
\begin{center}
\includegraphics{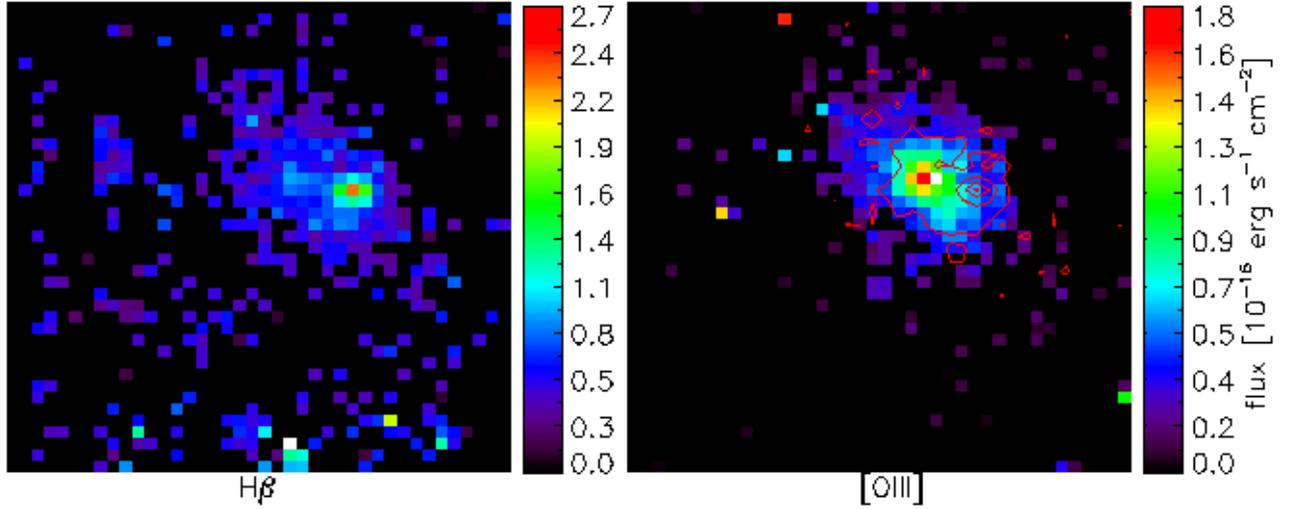}
\end{center}
\caption{\label{H_beta_OIII_linemap}$27"\times27"$ emission maps; north is up, east is left. Left: Line map of H$\beta$ emission. Right: Line map of [OIII] $\lambda 5008$ emission with overplotted contour lines for $95,\ 90,\ 80,\ 60\ $ and $40\%$ of the H$\beta$ emission.}
\end{figure}

\begin{figure}
\begin{center}
\includegraphics{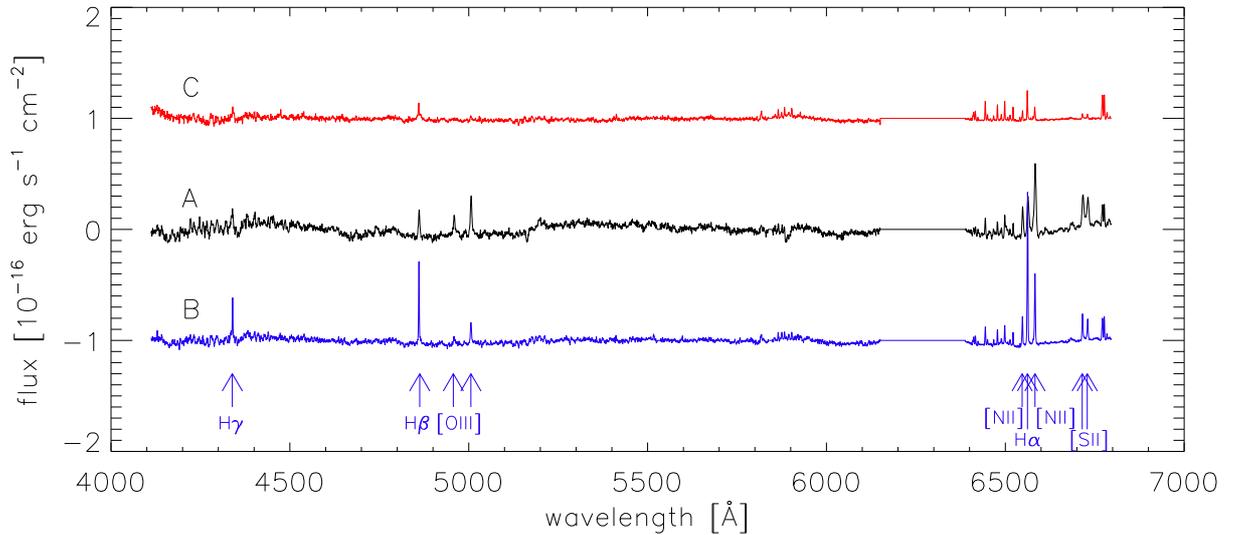}
\end{center}
\caption{\label{all_spectra}Spectra of the regions of interest stacked of 3 by 3 pixels centered on the respective brightest pixel. The spectra are denoted according to the regions depicted in Fig. \ref{H_alpha_NII_linemap}. For better clarity of the plot, the values of spectrum B and C were shifted by $\pm1\cdot10^{-16}\ \mathrm{erg\ sec}^{-1}\ \mathrm{cm}^{-2}$.}
\end{figure}

The H$\alpha$ emission line map in the left panel of Fig. \ref{H_alpha_NII_linemap} shows three peaks, one located on the continuum peak (A), a second one $0.46\ \mathrm{kpc}$ (B) west of the center and a third one east at $1.6\ \mathrm{kpc}$ (C) projected distance to the core (A). The comparison of the map with the one for [NII] $\lambda 6585$ line emission in the right panel of Fig. \ref{H_alpha_NII_linemap} implies, that location B and C are clearly distinctive in their ionisation mechanisms from the core, showing only little [NII] emission compared to the Balmer emission. This is also supported by the emission maps of H$\beta$ and [OIII] $\lambda 5008$ (left and right panel of Fig. \ref{H_beta_OIII_linemap}, respectively), where H$\beta$ is strong in region B, weak in the core region A and even weaker in region C. Unfortunately for region C no [OIII] fluxes could be obtained, because of low S/N. However, considering the distance of region C to the center, it is likely to be a star-forming region. Also to mention is the fact, that the [NII] $\lambda 6585$ emission does not appear to be confined to the core region but it extends to the south-west. In Fig. \ref{all_spectra} the stacked spectrum (consisting of 3 by 3 pixels) of each region marked in Fig. \ref{H_alpha_NII_linemap} is shown.\\

\begin{figure}
\begin{center}
\includegraphics{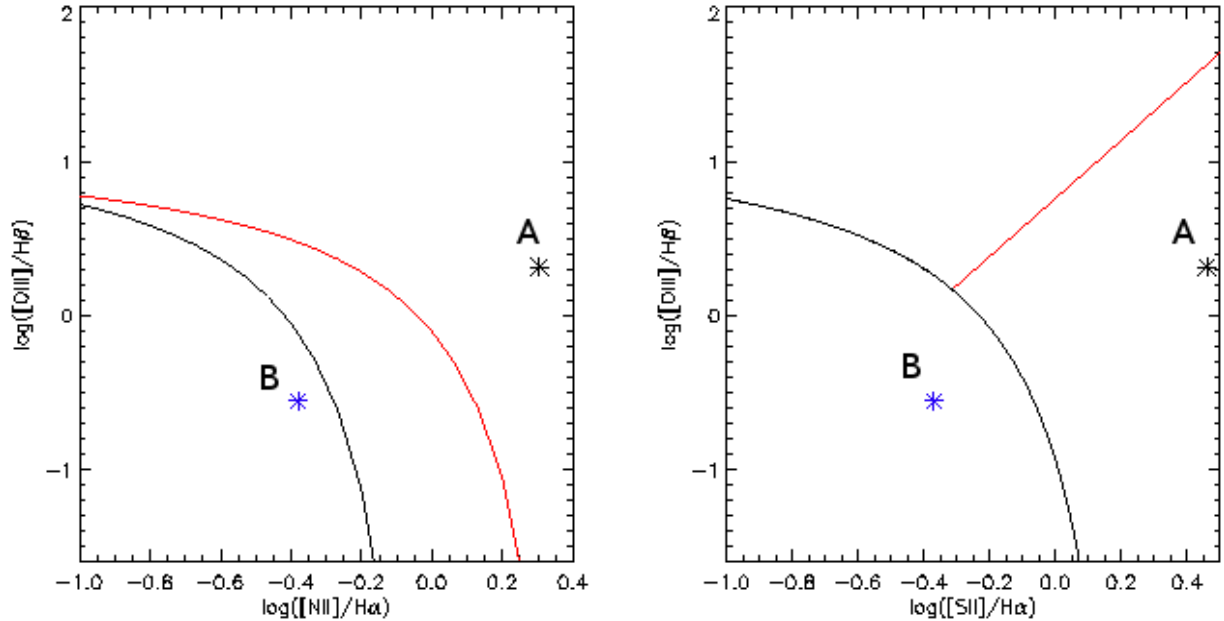}

\end{center}
\caption{\label{diagnostic}The locations of the brightest pixel in the regions A and B, as denoted in Fig. \ref{H_alpha_NII_linemap}, in the diagnostic diagrams. Left: The red line marks the maximum starburst line, that represents the border between the composite (HII + AGN) and the AGN domain \cite{Kewley_maximum_starburst}; the black line indicates the border between the pure star-forming region and the composite region \cite{Kauffmann_pure_SF}. Right: The black line indicates the maximum starburst line, the red line marks the border between the Seyfert and LINER classification \cite{Kewley_Sey_LINER}.}
\end{figure}

In agreement with \cite{Liner_reference}, the diagnostic diagram in Fig. \ref{diagnostic} reveals the brightest pixel of the core (A) to be a LINER. The brightest pixel in region B is classified as an HII region. Region C cannot be classified by the means of the diagnostic diagram, because its [OIII] $\lambda5008$ emission line could not be securely detected.

\subsection{Kinematics}

\begin{figure}
\begin{center}
\includegraphics{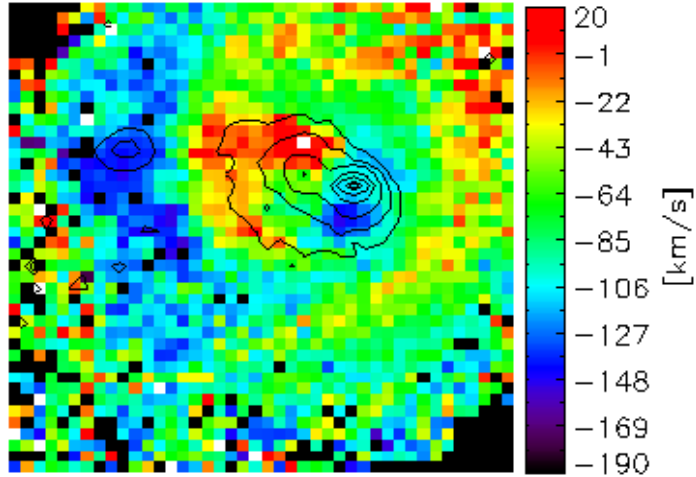}
\end{center}
\caption{\label{H_alpha_LOSV}The line of sight velocity field derived from the H$\alpha$ emission line with the same overplotted contours of H$\alpha$ emission as in the right panel of Fig. \ref{H_alpha_NII_linemap}. The field of view is $27"\times27"$; north is up, east is left.}
\end{figure}

The H$\alpha$ line of sight velocity map (Fig. \ref{H_alpha_LOSV}) shows a peculiar structure, resembling the typical 'Yin-Yang' pattern and is possibly dominated by the nuclear bar. The gas kinematics is completely different from the rotational stellar velocity field published by \cite{sigma_hollows_lorenzo}. The overplotted H$\alpha$ emission contours show the HII region (B) either being located in an area of close to zero velocity or the region's intrinsic velocity being lower than that of the surrounding gas. The core (A) is also in an area of relative rest, as expected, but it is also found close to a region of high redshifted velocities to the north and east. The star-forming region in the east (C) is associated with high blueshifted velocities. Though very noisy, the eastern edge of the field of view suggests the detection of the galaxy's inner ring.

\section{Conclusion and Outlook}

Our observations confirm the low activity of NGC 5850 and the emission line ratios put it in the LINER area of the diagnostic diagram. The maps also show two star forming regions (denoted as B and C) at $\sim1.6\ \mathrm{kpc}$ and $\sim0.46\ \mathrm{kpc}$ projected distance to the core (A). The HII-region (B) close to the core is an interesting object, in particular, regarding fuelling scenarios of the central engine and its feedback into the host galaxy. The complex gas kinematics introduced by the inner bar is likely to provide further insight into fuelling scenarios for NGC5850.\\

However, the question whether the core of a LINER is a downscaled AGN has not been answered, yet. The extended [NII] emission around the core suggests that the LINER emission might not only be caused by the unresolved central engine. Alternatively, the LINER emission could be caused by shock ionisation. Nevertheless, central point sources are capable of creating extended line emission \cite{Julia_EELR, Yan_nature_of_LINERS}, thus the presence of an AGN cannot be ruled out.

\ack
Part of this work was supported by the German Deutsche Forschungsgemeinschaft, DFG, via grant SFB 956 project A2 and fruitful discussions with members of the European Union funded COST Action MP0905: Black Holes in a violent Universe and PECS project No. 98040. M. B., J. S., J. Z. and S. F. gladly acknowledge the financial support by Group of Eight Australia-Germany Joint Research Co-operation Scheme via project-ID 50753527. M. V.-S. is member of the International Max Planck Research School (IMPRS) for Astronomy and Astrophysics at the MPIfR and the Universities of Bonn and Cologne.

\bibliographystyle{iopart-num}
\bibliography{js_jz_mvs_AHAR_NGC5850}

\end{document}